\documentclass[aps,floatfix]{revtex4}

\usepackage{amssymb,hyperref,hypcap,ae,tikz,color}
\usepackage{latexsym}
\usepackage{graphicx}
\usepackage{xcolor}

\usepackage[T1]{fontenc}
\usepackage[latin9]{inputenc}

\usepackage{mathrsfs}
\usepackage{amsmath}
\usepackage{amssymb}
\usepackage{esint}
\usepackage{babel}

\hypersetup{
	colorlinks=true,
	citecolor=green,
    linkcolor=magenta,
    urlcolor=blue
	}

\newcommand{\be}{\begin{equation}}
\newcommand{\ee}{\end{equation}}
\newcommand{\bea}{\begin{eqnarray}}
\newcommand{\eea}{\end{eqnarray}}

\begin{document}
\title{Formation of Singularity and Apparent Horizon for Dissipative Collapse in $f(R,T)$ Theory of Gravity }
\author{Uttaran Ghosh and Sarbari Guha}
\affiliation{Department of Physics, St. Xavier's College (Autonomous), Kolkata 700016, India}
\begin{abstract}
In this paper, we consider the spherically symmetric gravitational collapse of isotropic matter undergoing dissipation in the form of heat flux, with a generalized Vaidya exterior, in the context of $f(R, T)$ gravity. Choosing $f(R, T)=R+2\lambda T$, and applying the $f(R, T)$ junction conditions on the field equations for the interior and exterior regions, we have obtained matching conditions of the matter-Lagrangian and its derivatives across the boundary. The time of formation of singularity and the time of formation of apparent horizon have been determined and constraints on the integration constants are examined for which the final singularity is hidden behind the horizon.
\end{abstract}

\maketitle

\section{Introduction}
The study of gravitational collapse of a massive object has been an area of interest for many years. Depending on the mass of the collapsing object, the final stage of such collapse may lead to the formation of spacetime singularities where normal laws of physics are no longer valid. Over the years various aspects of collapse have been studied by several researchers \cite{Glass1981, Santos1985, Herrera_et_al2009, OlivSan1985, OliPachSantos1986, HerDenSan1989, ChanKicheDenSan1989, Chan1993, ChanHerSan1994, Chan2000, HerSanPRD04, HerSanDenGRG2012,OlivSan1987}, which include dynamical instability, causal heat transport, as well as the examination of the progress of collapse for different types of matter, which may or may not involve shear, heat flux, free-streaming radiation, and/or anisotropy of pressure. Different combinations of these factors, lead to the difference in the nature of the end state of collapse.
There have also been attempts to search for solutions to the Einstein Field Equations for various kinds of matter \cite{Chan1997, Chan2000}, where the Vaidya metric \cite{Vaidya1953} has been considered to describe the spacetime outside the collapsing matter.


The failure of General Relativity (GR) to account for the observed accelerated expansion of the universe as revealed from the  observational data from the Type Ia supernova \cite{SSTC, Perlmutter}, led scientists to propose an unknown component in the matter-energy sector of the universe, called the dark energy \cite{Felice}. In addition to this, the observation of the galactic rotation curves also indicate the existence of an entity called ``dark matter'', which seems to dominate the matter content of the universe. It seems that at extremely large scales, Einstein's theory of gravity is not a suitable option to explain the observed features of the universe. Hence, new theories of modified gravity were developed \cite{NojOdint2007, NojOdint2011, CapoLaurent2011, CliftFerrPadSkor2012, BertolamiSeqPRD2009, HarkoPRD2014, HarkoJCAP2014, SharifHEP2013} in order to explain astrophysical phenomena at large scales. One such theory was the $f(R)$ theory in which the Einstein-Hilbert action was generalised by replacing the Ricci scalar $R$ appearing in it, by a function of the Ricci scalar, $f(R)$, which depends only on the geometry of the spacetime and not on the matter content. This modification stemmed from the idea of including the higher order terms of curvature in the action integral, which give rise to the dark energy components. Moreover, the presence of these higher order curvature terms mimics the role of the cosmological constant $\Lambda$ in GR at the current epoch of evolution of the universe (thereby representing the $\Lambda$CDM model), and integrates all the phases of evolution of the universe in a single model \cite{CruzDombrizDobadoPRD2006, DunsbyPRD2010, NojiriOdintPRD2003}. Barraco and Hamity \cite{barraco2000} found spherically symmetric solutions in a first order approximation of $f(R)$ theory of gravity, and showed that both the exterior and interior metrics satisfy the junction conditions. They also showed that at least one exterior solution is the Schwarzschild metric. Capozziello, Stabile and Troisi \cite{Capozziello2007} used the existence of Noether symmetry to find spherically symmetric solutions in $f(R)$ theory. They also found classes of exact solutions for spherically symmetric case in $f(R)$ theory, both for constant Ricci scalar $R_{0}$ as well as $R(r)$ where $r$ is the radial coordinate \cite{Capozziello2008}. More spherically symmetric solutions in $f(R)$ gravity were found using this Noether symmetry approach \cite{Capozziello2012}. Multamaki and Vilja \cite{Multamaki2006} examined static empty space solutions with spherical symmetry in $f(R)$ theory of gravity. Chakrabarti and Banerjee \cite{ChakBanj2016} found the time of apparent horizon formation and singularity formation for a perfect fluid collapse in $f(R)$ gravity. Sharif and Kausar \cite{Sharif2010} found apparent horizons for spherically symmetric perfect fluid collapse in $f(R)$ theory.

The $f(R,T)$ theory, first proposed by Harko et al. \cite{Harko2011} is a further modification of GR, which can provide a suitable explanation for this accelerated expansion of the universe. This theory is a further generalisation of the $f(R)$ theory, in which, the Ricci scalar $R$ in the Einstein-Hilbert action is replaced by a function of $R$ and $T$, the latter being the trace of the energy-momentum tensor appearing in the Einstein field equations. The $f(R,T)$ function arises due to quantum effects or due to the existence of exotic imperfect matter fluids. In their paper \cite{Harko2011}, Harko and collaborators also discussed a few cases for possible choices of the $f(R,T)$ function. A suitable choice is the linear form $R+2\lambda T$ which gives rise to power-law type of scale factors in the corresponding cosmological model. Sahoo et al. \cite{Sahoo2018} showed that $f(R) + \lambda T$ gravity models act as alternatives to cosmic acceleration. The inclusion of the trace $T$ of the energy-momentum tensor in the Einstein-Hilbert action enables one to study the effect of curvature-matter interaction in the evolution of the universe \cite{Jamil2012,Sharif2013,Singh2014,Shabani2014,Noureen2015,Moraes2015,Sharif2021}.
Moraes and Sahoo constructed wormhole models \cite{MoraesSahoo2017a}, and also proposed a cosmological scenario from the simplest non-minimal matter-geometry coupling in the $f (R, T )$ theory of gravity \cite{MoraesSahoo2017b}. Moraes, Correa and Lobato \cite{Moraes2017} found solutions for a static wormhole metric in the $f(R,T)$ framework with linearised form of the $f(R,T)$ function. Sharif and Fatima \cite{Sharif2023} studied the the effects of charge on traversable wormhole
structure in $f(R, T)$ theory. Zaregonbadi, Farhoudi and Riazi \cite{zaregonbadi2016} found solutions for a static spherically symmetric spacetime in $f(R,T)$ gravity, and extracted the expressions for the metric components in the case of the galactic halo, using minimal coupling.
Amir and Sattar \cite{Amir2015} investigated spherically symmetric collapse of a perfect fluid in $f(R,T)$ gravity, and determined the conditions for the formation of the apparent horizon. Abbas and Ahmed \cite{Abbas2019} studied charged perfect fluid collapse and apparent horizon formation in $f(R,T)$ theory. Yousaf et al. \cite{Yousaf2016a, Yousaf2016b} studied the influence of structure scalars obtained by orthogonal splitting of the Riemann tensor on various physical properties of the collapsing matter, such as energy density inhomogeneity, pressure anisotropy and shear viscosity in $f(R,T)$ theory, and provided an insight on how the evolution of the collapsing matter proceeds under the effect of tidal forces and these inhomogeneities. Yousaf \cite{YousafPDU2020} considered modelling of a gravastar in cylindrical symmetry in $f(R,T)$ theory, which could be a suitable alternative to a black hole as the final state of the collapse, and studied the effect of electromagnetic field on the mass-energy content of the middle thin shell of the gravastar, and also drew comparisons between its density and pressure, and its proper length and thickness. In a previous work \cite{GuhaGhosh2021}, the present authors studied the dynamical stability and heat transport for a collapsing dissipative fluid in $f(R,T)$ gravity.

In this paper, we consider the region outside the collapsing matter to be described by the generalized Vaidya spacetime, which is then used to apply the $f(R,T)$ junction conditions \cite{RosaPRD2021} in order to solve the $f(R,T)$ field equations corresponding to the interior spacetime. We assume the geometry to be spherically symmetric so that there is no generation of gravitational waves, and the matter distribution involves pressure isotropy and heat flux. The spherical symmetry can also be used to model realistic gravitational collapse with small deviations only. The exterior region is assumed to be filled with a combination of Type-I and Type-II fluids, similar to what Wang and Wu \cite{WangGRG1999} had considered while working with a generalized Vaidya metric (which was a generalization of the original Vaidya metric \cite{Vaidya1951}). The metric coefficients are assumed to be separable into spatial and temporal parts, similar to the works of Sharif and Abbas \cite{SharifAbbasChargedCyl2011}, and Guha and Banerji \cite{GuhaBanerji2014}. The time of formation of the singularity, and that of the apparent horizon is determined and the nature of the resulting singularity is analysed in consideration of the work of Joshi, Goswami and Dadhich \cite{JoshiGosDadhPRD04}. The time of formation of apparent horizon for different types of collapse was studied previously in GR \cite{ChattGhoshJaryal,DebnathNathChak}. Here, a study of the same is carried out in $f(R,T)$ theory.

The organization of this paper is as follows: In Section II we provide a brief introduction to the $f(R,T)$ formalism, and proceed to consider a generalized Vaidya exterior spacetime and determine the field equations for a combination of type-I and type-II fluids in Section III. Section IV consists of the description of the interior spacetime and the field equations for the interior spacetime are determined. In Section V, we enlist the junction conditions for the $f(R,T)$ theory and examine the results that emerge after the application of the junction conditions. In Section VI, the time of formation of singularity is determined by examining the nature of the temporal dependence of the physical radius of the collapsing matter. Section VII deals with the formation of apparent horizon of the collapsing system, the nature of the final singularity, and the restrictions on the parameters governing the evolution of the system so as to ensure black hole formation. This is followed by a summarization and discussion in Section VIII.

\section{The $f(R,T)$ Formalism}

The modified Einstein-Hilbert action in $f(R,T)$ gravity, is given by
\begin{equation}\label{EHaction}
S=\int ~d^{4}x \sqrt{-g} \Bigg(\frac{f(R, T)}{16\pi G} + \mathcal{L} _ {m} \Bigg),
\end{equation}
where $g$ is the determinant of the spacetime metric, $G$ is the gravitational constant, $f(R,T)$ is an arbitrary function of the Ricci scalar $R$ and the trace $T$ of the energy-momentum tensor of the matter filling up the spacetime, and $L_{m}$ is the matter Lagrangian.
The field equations of $f(R,T)$ gravity \cite{Harko2011} can be written in the form :
\begin{equation}\label{fRTfe}
R_{\mu\nu}=\frac{1}{f_{R}}\left[\left(1+f_{T}\right)T^{m}_{\mu\nu}-L_{m}g_{\mu\nu}f_{T} +\frac{1}{2}g_{\mu\nu}f\textcolor{red}{-}D_{\mu\nu}\right],
\end{equation}
where $g_{\mu\nu}$ is the spacetime metric, $R$ is the Ricci scalar, $T$ is the trace of the energy-momentum
tensor, $f_{R}$ and $f_{T}$ are the derivatives of the $f(R,T)$
function with respect to $R$ and $T$, respectively, $L_{m}$ is
the matter Lagrangian, and $D_{\mu\nu}=\left(g_{\mu\nu}\Box-\nabla_{\mu}\nabla_{\nu}\right)f_{R}$
which includes the higher order curvature terms, and acts as the
source of dark energy.

In order to obtain the form of the field equation \eqref{fRTfe} in $f(R,T)$ gravity, it was assumed by Harko in \cite{Harko2011} that the matter Lagrangian was minimally coupled to the metric, that is, it depended only on the components of the metric tensor $g_{\mu\nu}$ but not on the derivatives of $g_{\mu\nu}$.
In what follows, we will choose $f(R,T)=R+2\lambda T$ for our analysis.

\section{The Exterior Spacetime}
The boundary of the collapsing matter divides the entire spacetime into an interior region containing the collapsing matter, and an exterior region containing the material particles, photons, and thermal radiation emanating from the star during the collapse. The exterior spacetime in the region outside the boundary of the collapsing body is considered to be described by the generalised Vaidya metric of outgoing radiation, and is given by
\begin{equation}\label{genvaidya}
ds_{+}^{2}=-\left(1-\frac{2M(v,Y)}{Y}\right)dv^{2}-2dvdY+Y^{2}d\Omega^{2},
\end{equation}
where $d\Omega^{2}=d\theta^{2}+\sin^{2}\theta d\phi^{2}$. Here, $v$ is the retarded null coordinate, and $Y$ is the radial coordinate for the exterior region. The parameter $M(v,Y)$ represents the mass-energy content inside a radius $Y$ at a time $v$.

The exterior energy-momentum tensor is assumed to represent a combination of type-I and type-II fluids \cite{WangGRG1999}, and is expressed as follows:
\begin{equation}
T_{\mu\nu}^{+}=\mu l_{\mu}l_{\nu}+\left(\rho+P\right)\left(l_{\mu}n_{\nu}+l_{\nu}n_{\mu}\right)+Pg_{\mu\nu}^{+},
\end{equation}
where the first term represents a type-I fluid of density $\mu$, which represents null-like matter/radiation photons, and the second and third terms together represent a type-II fluid, with $\rho$ being its density and $P$ being its isotropic pressure, which represents a perfect fluid component which can be timelike/massive particles. For a type-I fluid, the energy-momentum tensor has one timelike eigenvector, while for a type-II fluid, ithe energy-momentum tensor has a double null eigenvector. A discussion on various types of matters can be found in the book \cite{HawkingEllis}.
Here, $g_{\mu\nu}^{+}$ is the metric tensor for the exterior spacetime. The positive sign in the subscript of the line element and the superscript of the energy-momentum tensor denotes the exterior spacetime. Further,
\begin{equation}
l_{\mu}=\delta_{\mu}^{0},
\end{equation}
\begin{equation}
\textrm{and} \qquad n_{\mu}=\frac{1}{2}\left(1-\frac{2M(v,Y)}{Y}\right)\delta_{\mu}^{0}-\delta_{\mu}^{1} .
\end{equation}

The trace of the exterior energy-momentum tensor is
\begin{equation}\label{Traceexterior}
T^{+}=6P+2\rho .
\end{equation}
Invoking the equation of state for pure radiation, $P=\rho/3$, and applying it to equation \eqref{Traceexterior}, we obtain $T^{+}=4\rho$.

On account of our choice $f(R,T)=R+2\lambda T$, we find that $f_{R}=1$, $f_{T}=2\lambda$ and $D_{\mu\nu}=0$.
In that case, the $f(R,T)$ field equations for the exterior spacetime reduces to :
\begin{align}
R_{00} & =\frac{1}{Y^{2}}\left(\left(\frac{\partial^{2}M}{\partial Y^{2}}\right)\left(2M-Y\right)-2\frac{\partial M}{\partial v}\right) =\left(1-\frac{2M}{Y}\right)\left(\left(1+2\lambda\right)\rho+2\lambda L_{m_{ext}}-\frac{f}{2}\right)+\left(1+2\lambda\right)\mu,\\
R_{11} & =0,\\
R_{22} & =2\frac{\partial M}{\partial Y}=Y^{2}\left[\left(1+2\lambda\right)P-2\lambda L_{m_{ext}}+\frac{f}{2}\right],\\
R_{33} & =R_{22}\sin^{2}\theta,\\
R_{01} & =-\frac{1}{Y}\frac{\partial^{2}M}{\partial Y^{2}}=-\left(1+2\lambda\right)\left(\rho+P\right)-\frac{R_{22}}{Y^{2}}.
\end{align}
where, $L_{m_{ext}}$ is the matter Lagrangian for the exterior spacetime.

\section{The Interior Spacetime}
The interior spacetime is assumed to be of the most general spherically symmetric form, given by
\begin{equation}\label{intmetric}
ds_{-}^{2}=-A(r,t)^{2}dt^{2}+B(r,t)^{2}dr^{2}+C(r,t)^{2}\left(d\theta^{2}+\sin^{2}\theta d\phi^{2}\right).
\end{equation}

The matter-energy momentum tensor for the interior spacetime is considered
to be that of an isotropic fluid undergoing dissipation in the form
of heat flux, and is given by
\begin{equation}\label{emtinterior}
T_{\mu\nu}^{(m)^{-}}=\left(\rho_{int}+p\right)u_{\mu}u_{\nu}+pg_{\mu\nu}^{-}+q_{\mu}u_{\nu}+q_{\nu}u_{\mu},
\end{equation}
where, $\rho_{int}$ is the interior energy density, $p$ is the isotropic pressure, and the four-velocity is given by $u^{\mu}=A^{-1}\delta_{0}^{\mu}$. We define $\chi^{\mu}=B^{-1}\delta_{1}^{\mu}$, where $\chi^{\mu}$ is a unit vector in the radial direction. Then the heat flux vector is defined as
$q^{\mu}=q\chi^{\mu}$ where $q=q(r,t)$ is the heat flux, and the four-velocity satisfies
$u^{\mu}u_{\mu}=-1$.
The trace of the interior energy-momentum tensor is given by
\begin{equation}\label{traceinterior}
T^{-}=-\rho_{int}+3p.
\end{equation}
The $f(R,T)$ field equations are given by
\begin{equation}\label{fRTfeInt}
R_{\mu\nu}=\frac{1}{f_{R}}\left[\left(1+f_{T}\right)T^{m^{-}}_{\mu\nu}-L_{m}g_{\mu\nu}^{-}f_{T} +\frac{1}{2}g_{\mu\nu}^{-}f-D_{\mu\nu}\right],
\end{equation}
where $R_{\mu\nu}$ is the Ricci tensor, $R$ is the Ricci scalar, $T$ is the trace of the energy-momentum
tensor, $f_{R}$ and $f_{T}$ are the derivatives of the $f(R,T)$
function with respect to $R$ and $T$, respectively, $L_{m}$ is
the interior matter Lagrangian, and $D_{\mu\nu}=\left(g_{\mu\nu}\Box-\nabla_{\mu}\nabla_{\nu}\right)f_{R}$
includes the higher order curvature terms, which acts as the
source of dark energy.

Just as in the case of the exterior spacetime, the choice $f(R,T)=R+2\lambda T$, implies that $f_{R}=1$, $f_{T}=2\lambda$
and $D_{\mu\nu}=0$. Evidently, the Ricci scalar $R$ and the trace $T$ will be different for the interior and the exterior spacetimes. The interior field equations then take the following
form :
\begin{align}\label{field00}
R_{00}=-\frac{2\ddot{C}}{C}+\frac{2\dot{A}\dot{C}}{AC}-\frac{\ddot{B}}{B}+\frac{\dot{A}\dot{B}}{AB} +\frac{2AA'C'}{B^{2}C}+\frac{AA''}{B^{2}}-\frac{AA'B'}{B^{3}} & =A^{2}\left[\left(1+2\lambda\right)\rho_{int}+2\lambda L_{m}-\frac{f}{2}\right], \\ \label{field11}
R_{11}=-\frac{2C''}{C}+\frac{B\ddot{B}}{A^{2}}+\frac{2B\dot{B}\dot{C}}{A^{2}C}-\frac{\dot{A}\dot{B}B}{A^{3}} -\frac{A''}{A}+\frac{2B'C'}{BC}+\frac{A'B'}{AB} & =B^{2}\left(\left(1+2\lambda\right)p-2\lambda L_{m}+\frac{f}{2}\right), \\ \label{field22}
R_{22}=-\frac{CC''}{B^{2}}+\frac{C\ddot{C}}{A^{2}}-\frac{C\dot{C}\dot{A}}{A^{3}}+\frac{C\dot{C}\dot{B}}{A^{2}B} -\frac{CC'A'}{AB^{2}}+\frac{CC'B'}{B^{3}}+\frac{2m(r,t)}{C} & =C^{2}\left[\left(1+2\lambda\right)p-2\lambda L_{m}+\frac{f}{2}\right], \\ \label{field33}
R_{33} & =R_{22}\sin^{2}\theta, \\ \label{field01}
R_{01}=-\frac{2\dot{C}'}{C}+\frac{2A'\dot{C}}{AC}+\frac{2\dot{B}C'}{BC} & =-\left(1+2\lambda\right)qAB
\end{align}
where
\begin{equation}
m(r,t)=\frac{C}{2}\left(1+\frac{\dot{C}^{2}}{A^{2}}-\frac{C'^{2}}{B^{2}}\right).
\end{equation}

\section{The Junction Conditions}

The junction conditions represent the continuity of relevant quantities across the boundary of the collapsing star. In their works, Darmois and Israel had presented and discussed the junction conditions in the context of GR \cite{Darmois, Israel}. The continuity of the line element and the extrinsic curvature tensor are the only necessary junction conditions in GR. The consideration of junction conditions depends on the nature of the Einstein-Hilbert action in the various theories of gravity. In the $f(R)$ theory of gravity, in addition to the junction conditions in GR, one has to include the continuation of the Ricci scalar and its derivative. The continuity of the trace part and the trace-free parts of the extrinsic curvature tensor are also to be considered separately. Junction conditions in $f(R)$ gravity were discussed by the authors in \cite{Deruelle2008, CliftonPRD2013, SenovillaPRD2013, GoswamiPRD2014}. Further, in the $f(R,T)$ theory, the trace of the energy-momentum tensor and the derivative of the trace are two additional quantities required to be matched across the hypersurface $\Sigma$.

The timelike 3D-hypersurface $\Sigma$ which separates the interior and the exterior spacetime, and serves as the boundary of the collapsing matter, is given by
\begin{equation}\label{hypersurface}
ds_{\Sigma}^{2}=-d\tau^{2}+\mathcal{R}(\tau)^{2}\left(d\theta^{2}+\sin^{2}\theta d\phi^{2}\right),
\end{equation}
where $(\tau,\theta,\phi)$ represents the coordinates on the hypersurface $\Sigma$, $\mathcal{R}(\tau)$ is the radius of the 2-sphere specified by the angular coordinates.

The junction conditions for the perfect-fluid version of $f(R,T)$ gravity (both in the geometrical representation as well as in a dynamically equivalent scalar-tensor representation) were derived by Rosa \cite{RosaPRD2021}. In the particular case of a smooth matching with no thin-shell, those junction
conditions are:
\begin{align}
\label{jc1}\left[g_{\mu\nu}\right]_{-}^{+} & =0,\\
\label{jc2}\left[\tilde{K}_{ij}\right]_{-}^{+} & =0,\\
\label{jc3}\left[K\right]_{-}^{+} & =0,\\
\label{jc4}\left[R\right]_{-}^{+} & =0,\\
\label{jc5}\left[T\right]_{-}^{+} & =0,\\
\label{jc6}\left[\partial_{\mu}T\right]_{-}^{+} & =0,\\
\label{jc7}\left[\partial_{\mu}R\right]_{-}^{+} & =0.
\end{align}
Here, the Greek indices represent the coordinates of the 4D-spacetimes,
while the Latin indices represent the coordinates of the hypersurface.
The negative sign signifies the interior spacetime, while the positive
sign signifies the exterior spacetime.

The junction condition in \eqref{jc1}, which is similar to the first Darmois-Israel junction condition \cite{Darmois, Israel} can be written as
\begin{equation}
ds_{-}^{2}=ds_{\Sigma}^{2}=ds_{+}^{2},
\end{equation}
which gives us
\begin{align}\label{eq1jc1}
A^{2}\left(r,t\right)dt^{2}\left|_{\Sigma}\right. & =d\tau^{2}\left|_{\Sigma}\right.=\left(1-\frac{2M(v,Y)}{Y}+2\frac{dY}{dv}\right)dv^{2}\left|_{\Sigma}\right.,\\
\label{eq2jc1} \quad \textrm{and} \qquad C\left(t,r_{\Sigma}\right) & =\mathcal{R}_{\Sigma}\left(\tau\right)=Y_{\Sigma}\left(\tau\right).
\end{align}

The junction conditions in \eqref{jc2} and \eqref{jc3} matches the
trace-free parts and the trace parts of the extrinsic curvature tensors
for the two spacetime regions respectively, across the bounding surface.
The expression for the extrinsic curvature tensor is given by
\begin{equation}\label{kij}
K_{ij}=-N_{\sigma}\left(\frac{\partial^{2}\psi^{\sigma}}{\partial\xi^{i}\partial\xi^{j}} +\Gamma_{\alpha\beta}^{\sigma}\frac{\partial\psi^{\alpha}}{\partial\xi^{i}}\frac{\partial\psi^{\beta}}{\partial\xi^{j}}\right),
\end{equation}
where $\xi^{i} \equiv (\tau,\theta,\phi)$ are the coordinates on the hypersurface, $\Gamma_{\alpha\beta}^{\sigma}$ are the Christoffel symbols for the spacetime whose extrinsic curvature components are being evaluated, $N_{\sigma}$ is the normal to the hypersurface, and $\psi^{\sigma}$ are the coordinates of the 4D-spacetime.

The trace of the extrinsic curvature tensor is given by
\begin{equation}\label{ktrace}
K=h^{ij}K_{ij} ,
\end{equation}
where $h_{ij}$ is the metric on the hypersurface.
The traceless part of the extrinsic curvature tensor is given by
\begin{equation}\label{ktraceless}
\tilde{K}_{ij}=K_{ij}-\frac{1}{3}h_{ij}K.
\end{equation}
Matching the quantities $K$ and $\tilde{K}_{ij}$ across the boundary of the collapsing matter, we have
\begin{align}
\left[-K_{\tau\tau}^{-}+\frac{2}{\mathcal{R}}K_{\theta\theta}^{-}\right]_{\Sigma} & =\left[-K_{\tau\tau}^{+}+\frac{2}{\mathcal{R}}K_{\theta\theta}^{+}\right]_{\Sigma},\\
\left[K_{\tau\tau}^{-}+\frac{1}{\mathcal{R}^{2}}K_{\theta\theta}^{-}\right]_{\Sigma} & =\left[K_{\tau\tau}^{+}+\frac{1}{\mathcal{R}^{2}}K_{\theta\theta}^{+}\right]_{\Sigma},
\end{align}
which when combined together, gives us the condition
\begin{equation}\label{extcurvmatch}
\left(K_{ij}^{-}\right)_{\Sigma}=\left(K_{ij}^{+}\right)_{\Sigma},
\end{equation}
which is the same as the second Darmois-Israel junction condition \cite{Darmois, Israel}.
The condition \eqref{extcurvmatch} leads us to the following two equations :
\begin{align}
-\left(\frac{A'}{AB}\right)_{\Sigma} & =\left[\left(\frac{dv}{d\tau}\right)^{-1}\left(\frac{d^{2}v}{d\tau^{2}}\right) -\frac{1}{Y}\left(\frac{dv}{d\tau}\right)\left(\frac{M}{Y}-\frac{dM}{dY}\right)\right]_{\Sigma},\\
\left(\frac{CC'}{B}\right)_{\Sigma} & =\left[Y\left(\frac{dv}{d\tau}\right)\left(\left(\frac{dv}{d\tau}\right)^{-2} -\left(\frac{dY}{d\tau}\right)\left(\frac{dv}{d\tau}\right)^{-1}\right)\right]_{\Sigma}.
\end{align}
The junction conditions \eqref{jc1}, \eqref{jc2} and \eqref{jc3} deal with terms that are purely geometrical in their origin. In order to examine the rest of the junction conditions, which involve the continuity of the Ricci scalars, the traces, and their derivatives across the boundary, we now need to look into the nature of the energy-momentum tensor in the interior spacetime, since the Ricci scalar can be expressed in terms of the matter-energy components with the help of the field equations, and the trace originates from the energy-momentum tensor.

Using the junction conditions \eqref{jc1}, \eqref{jc2} and \eqref{jc3}, along with \eqref{field11} and \eqref{field01}, and noting that $m(r,t)$ is the total mass-energy content inside the collapsing matter at a time $t$ and radius $r$, which is the same as $M(v,Y)$ in \eqref{genvaidya}, we obtain
\begin{equation}\label{pqrel}
-\frac{C}{2}\left[\left(1+2\lambda\right)\left(p-q\right)-2\lambda L_{m}+\lambda T\right]\mid_{\Sigma}=\frac{1}{Y}\frac{dM}{dY}\mid_{\Sigma}.
\end{equation}
However, $m$ and $M$ have different functional dependence, and $\frac{dm}{dr}$ and $\frac{dM}{dY}$ are not necessarily equal, because these variations are not similar in general. From equation \eqref{pqrel}, it is evident that the pressure is related to the heat flux.

The junction conditions \eqref{jc4} and \eqref{jc5} implies that
\begin{equation}\label{JC_frt}
f(R,T)^{+}=f(R,T)^{-}.
\end{equation}
Further, \eqref{jc4} also gives us
\begin{equation}\label{JC_Ricciscalar}
R^{-}\mid_{\Sigma}=R^{+}\mid_{\Sigma},
\end{equation}
which when written in terms of the interior and the exterior field
equations, yields
\begin{equation}
\left(1+2\lambda\right)\left(-\rho_{int}+3p\right)-8\lambda L_{m}+2f^{-}\mid_{\Sigma}=\left(1+2\lambda\right)\left(2\rho+6P\right)-8\lambda L_{m_{ext}}+2f^{+}\mid_{\Sigma},
\end{equation}
while the condition \eqref{jc5} gives us
\begin{equation}\label{JC_trace}
-\rho_{int}+3p=2\rho+6P.
\end{equation}

Combining the four equations \eqref{JC_frt} to \eqref{JC_trace}, we get
\begin{equation}
-8\lambda L_{m}\mid_{\Sigma}=-8\lambda L_{m_{ext}}\mid_{\Sigma}.
\end{equation}

For $\lambda=0$, the $f(R,T)$ theory reduces to the case of Einstein's General Relativity. Hence, $\lambda$ must be non-vanishing for $f(R,T)$ theory. For the $f(R,T)$ junction conditions to be satisfied, the interior matter Lagrangian must be equal to the exterior matter Lagrangian at the boundary of the collapsing matter. Hence the $f(R,T)$ junction conditions impose a restriction on the choice of the matter Lagrangian. This is evident since the matter-Lagrangians for both the interior and the exterior spacetimes were considered to be the same while the junction conditions were formulated in \cite{RosaPRD2021}. If there is minimal coupling between the collapsing matter and the gravity, the interior matter Lagrangian can be chosen either as the interior matter-energy density $-\rho_{int}$, or the isotropic pressure of the matter, i.e. $p$ \cite{Faraoni2009, GuhaGhosh2021}. This freedom of choice is no longer available if the pressure of the interior matter is anisotropic. In this work, we have chosen the interior matter Lagrangian to be given by the interior energy density $\rho_{int}$.

The junction condition \eqref{jc6} then gives us
\begin{align}
-\frac{\partial\rho_{int}}{\partial t}+3\frac{\partial p}{\partial t} & \mid_{\Sigma}=2\frac{\partial\rho}{\partial v}+6\frac{\partial P}{\partial v}\mid_{\Sigma},\\
-\frac{\partial\rho_{int}}{\partial r}+3\frac{\partial p}{\partial r} & \mid_{\Sigma}=2\frac{\partial\rho}{\partial Y}+6\frac{\partial P}{\partial Y}\mid_{\Sigma}.
\end{align}

In order to examine the final junction condition \eqref{jc7}, the Ricci scalars
of the two spacetimes are expressed in terms of the respective field
equations, so that \eqref{jc7} leads us to the two equations
\begin{align}
\left(1+2\lambda\right)\left(-\frac{\partial\rho_{int}}{\partial t}+3\frac{\partial p}{\partial t}\right)-8\lambda \frac{\partial L_{m}}{\partial t}+2\frac{df(R,T)^{-}}{dt}\mid_{\Sigma} & =\left(1+2\lambda\right)\left(2\frac{\partial\rho}{\partial v}+\textcolor{blue}{6}\frac{\partial P}{\partial v}\right)-8\lambda \frac{\partial L_{m_{ext}}}{\partial v}+2\frac{df(R,T)^{+}}{dv}\mid_{\Sigma},\\
\left(1+2\lambda\right)\left(-\frac{\partial\rho_{int}}{\partial r}+3\frac{\partial p}{\partial r}\right)-8\lambda \frac{\partial L_{m}}{\partial r}+2\frac{df(R,T)^{-}}{dr}\mid_{\Sigma} & =\left(1+2\lambda\right)\left(2\frac{\partial\rho}{\partial Y}+\textcolor{blue}{6}\frac{\partial P}{\partial Y}\right)-8\lambda \frac{\partial L_{m_{ext}}}{\partial Y}+2\frac{df(R,T)^{+}}{dY}\mid_{\Sigma}.
\end{align}

In view of the junction condition \eqref{jc6}, we see that the bracketed terms
in each of the above two equations cancel out on either sides. Using $f^{-}=R^{-}+2\lambda T^{-}$ and $f^{+}=R^{+}+2\lambda T^{+}$, we arrive at the following relation :
\begin{equation}
\left[\partial_{\mu}L_{m}\right]_{-}^{+}=0.
\end{equation}
Hence, like the Ricci scalar and the trace and their derivatives, the matter Lagrangian and its derivative for the two spacetimes also need to be continuously matched across the boundary.

\section{Formation of Singularity}
We now proceed to examine the nature of the end state of gravitational collapse. As the particles in the collapsing ball get closer to each another in course of time, eventually a state of extreme high density and curvature will be reached, where there is a breakdown of the normal laws of physics, leading to a `singularity'. It is worthwhile to determine the time at which the singularity formation will take place, and whether the singularity will be a black hole, or a `naked singularity' (which is visible to the external observer). For this purpose, the temporal dependence of the physical radius of the collapsing star have to be examined in detail.

In view of the choice for the interior matter Lagrangian as $-\rho_{int}$ , the interior field equations now reduce to
\begin{align}
\label{fe1}
\frac{A''}{AB^{2}}-\frac{1}{A^{2}}\left(\frac{\ddot{B}}{B}+\frac{2\ddot{C}}{C}\right) +\frac{\dot{A}}{A^{3}}\left(\frac{\dot{B}}{B}+\frac{2\dot{C}}{C}\right) -\frac{A'B'}{AB^{3}}+\frac{2A'C'}{AB^{2}C} & =\rho_{int}-\frac{R}{2}-\lambda T, \\
\label{fe2}
-\frac{A''}{AB^{2}}-\frac{2C''}{B^{2}C}+\frac{\ddot{B}}{A^{2}B}-\frac{\dot{A}\dot{B}}{A^{3}B} +\frac{2\dot{B}\dot{C}}{A^{2}BC}+\frac{A'B'}{AB^{3}} +\frac{2B'C'}{B^{3}C} & =p+2\lambda\left(p+\rho_{int}\right)+\frac{R}{2}+\lambda T, \\
\label{fe3}
\frac{1}{C^{2}}\left(1+\frac{\dot{C}^{2}}{A^{2}}-\frac{C'^{2}}{B^{2}}\right) +\frac{C}{A^{2}}\left(\frac{\ddot{C}}{C^{2}}+\frac{\dot{C}\dot{B}}{C^{2}B} -\frac{\dot{C}\dot{A}}{C^{2}A}\right)-\frac{C''}{B^{2}C}+\frac{B'C'}{B^{3}C}-\frac{A'C'}{AB^{2}C} & =p+2\lambda\left(p+\rho_{int}\right)+\frac{R}{2}+\lambda T ,\\
\label{fe4}
\frac{\dot{C}'}{C}-\frac{A'\dot{C}}{AC}-\frac{\dot{B}C'}{BC} & =\left(1+2\lambda\right)\frac{qAB}{2}.
\end{align}

On account of isotropy of pressure, equations \eqref{fe2} and \eqref{fe3} yield us
\begin{align}\label{pressureisotropy}
-\frac{1}{AB^{2}}\left(A''-\frac{A'B'}{B}-\frac{A'C'}{C}\right)+\frac{1}{A^{2}B}\left(\ddot{B} -\frac{\dot{A}\dot{B}}{A}+\frac{\dot{B}\dot{C}}{C}\right)-\frac{1}{B^{2}C}\left(C''-\frac{B'C'}{B}\right)\nonumber \\
-\frac{1}{A^{2}C}\left(\ddot{C}-\frac{\dot{A}\dot{C}}{A}\right)-\frac{1}{C^{2}}\left(1+\frac{\dot{C}^{2}}{A^{2}} -\frac{C'^{2}}{B^{2}}\right) & =0.
\end{align}

Let us assume that the geometry of the dissipative collapse is described by the following profile \cite{GuhaBanerji2014}:
\begin{align}
A(r,t) & =A_{0}(r)s_{1}(t),\\
B(r,t) & =B_{0}(r)s_{2}(t),\\
C(r,t) & =rB_{0}(r)s_{3}(t).
\end{align}

Here, $A_{0}(r)$ and $B_{0}(r)$ represent the static perfect fluid configurations
just before the collapse begins. With the onset of collapse at $t=0$, the collapsing ball behaves as a non-adiabatic fluid undergoing dissipation
in the form of thermal radiation, and the dissipative effects are dependent on time.
This time-dependence is encoded in the functions $s_{1}(t)$, $s_{2}(t)$ and $s_{3}(t)$, which, when subjected to appropriate rescaling of the coordinate time, as in \cite{Chan1997, Chan2000,GuhaBanerji2014}, can be replaced by a single temporal function $w(t)$ associated with the cross-section of the collapsing fluid normal to the radial direction. A way of rescaling is to consider $s_{1}(t)=s_{2}(t)$, and $w(t)=\frac{s_{3}(t)}{s_{1}(t)}$, and make the transformations $s_{i}(t) \rightarrow \frac{s_{i}(t)}{s_{1}(t)}$ where $i=1,2,3$ .  Metric
coefficients separable in spatial and temporal parts, have been previously
considered for dissipative collapse of charged cylindrical anisotropic fluid in GR by Guha and Banerjee \cite{GuhaBanerji2014}.

Therefore, we can write
\begin{align}\label{A}
A(r,t) & =A_{0}(r), \\
\label{B} B(r,t) & =B_{0}(r), \\
\label{C} C(r,t) & =rB_{0}(r)w(t).
\end{align}

It may be mentioned here that the assumption of the same time dependence for all three metric coefficients leads us to a time independent equation of pressure isotropy, which cannot be used to find the form of the function $w(t)$. Hence such an assumption is not useful in this case.

Substituting the expressions for the metric coefficients defined in \eqref{A},\eqref{B} and \eqref{C} in the equation \eqref{pressureisotropy}, we get
\begin{equation}\label{diffeqn}
\frac{1}{A_{0}^{2}}\left(\frac{\ddot{w}}{w}+\frac{\dot{w}^{2}}{w^{2}}\right)+\frac{1}{r^{2}B_{0}^{2}w^{2}}-D(r)=0,
\end{equation}
where
\begin{equation}\label{Dr}
D(r)=-\frac{1}{A_{0}B_{0}^{2}}\left(A_{0}''-\frac{2A_{0}'B_{0}'}{B_{0}}-\frac{A_{0}'}{r}\right)-\frac{B_{0}''}{B_{0}^{3}} +\frac{B_{0}'}{rB_{0}^{3}}+\frac{2B_{0}'^{2}}{B_{0}^{4}}+\frac{1}{r^{2}B_{0}^{2}}.
\end{equation}
The function $D(r)$ contains all the radial derivatives. From a single differential equation containing radial derivatives of both $A_{0}(r)$ and $B_{0}(r)$, it is not possible to find individual functional forms for either $A_{0}(r)$ or $B_{0}(r)$. However, since we are interested in determining the time of formation of singularity at the end of collapse, it is useful to examine the time evolution of the radius of the collapsing matter, for which we aim to solve this second order differential equation \eqref{diffeqn} for the temporal part $w(t)$ of the sectional radius $C(r,t)$. Once the form of $w(t)$ is determined, the time of formation of singularity can be obtained from its functional form. In that case, a non-vanishing $D(r)$ will yield a different form of $w(t)$ compared to the case when $D(r)=0$ .

We examine the cases for which $D(r)\ne 0$, and $D(r)=0$, separately.

\subsection{Case I : $D(r)\ne0$}\label{CaseI}

This is the more complicated case and can only be analysed logically, as is outlined below.

We solve equation \eqref{diffeqn} for $w$ with the help of GRTensor software package, to obtain
\begin{equation}\label{w}
w=\pm\frac{1}{\sqrt{2Dk^{3}}}\left(C_{1}Hck^{2}-2dk-C_{2}Hc\right)^{1/2},
\end{equation}
where $C_{1}$ and $C_{2}$ are integration constants, and
\begin{align}\label{cdefine}
c(r)=\frac{1}{A_{0}^{2}},\\
\label{ddefine} d(r)=\frac{1}{r^{2}B_{0}^{2}}, \\
\label{Hdefine} H(r)=\sqrt{\frac{2D}{c}}, \\
\label{kdefine}
k=\textrm{exp}(Ht).
\end{align}
Since $w$ denotes the temporal dependence of the sectional radius which cannot be negative, we consider only the positive solution. For collapse to progress, we require $\dot{w}<0$.
A singularity will occur at the end state of collapse when the star will collapse to zero volume, as $w(t) \rightarrow 0$. This is where the singularity occurs at the final state of the collapse, which is different from the regular center of the star at $r=0$ before the formation of singularity where the values of the physical parameters like the star density and the curvature scalars do not diverge \cite{JoshiGravColandSptimeSing2007}.

The situation for which $w(t)=0$, provides us the time of formation of the singularity ($t_{s}$):
\begin{equation}\label{t}
t_{s}=\frac{1}{H}\ln\left(\frac{d\pm\sqrt{d^{2}+C_{1}C_{2}\left(Hc\right)^{2}}}{C_{1}Hc}\right).
\end{equation}

For real values of the discriminant in \eqref{t}, we get the condition
\begin{equation}\label{real_discr}
-\frac{d^{2}}{\left(Hc\right)^{2}}\le C_{1}C_{2},
\end{equation}
and additionally, the argument of the logarithm needs to be positive.

As detailed in Appendix A, the parameters $H$ and $\frac{d}{c}$ must be constants. Therefore, $k$ is independent of $r$. These constants are now denoted by the following symbols :

\begin{align}\label{Hconstant}
H & =C_{3}, \\
\label{C4define} \frac{d}{c} & =C_{4}.
\end{align}

If we examine equation \eqref{C4define} closely, we see that on account of \eqref{cdefine} and \eqref{ddefine}, this ratio actually implies the condition
\begin{equation}\label{ABrelation}
A_{0}^{2}=C_{4}r^{2}B_{0}^{2}.
\end{equation}
To understand the physical significance of this condition, we consider the radial null geodesics, which, for our interior line element are given by $ds^{2}=0$, with $\theta$ and $\phi$ as constant.
In this case, using \eqref{ABrelation}, we find (ignoring the constant),
\begin{equation}\label{dtdr}
\frac{dt}{dr}=\pm\frac{1}{r},
\end{equation}
or,
\begin{equation}\label{trrelation}
t=\pm ln\left|r\right|.
\end{equation}
The slope of the curve tends to infinity as one approaches the regular center of the collapsing star at $r=0$, and the light cones tend to close up, in the vicinity of that point. This means that a signal takes infinite time to reach an external observer from that point, and the singularity at that point is a black hole. The singularity at the center of the collapsing matter at $r=0$ is not a removable singularity which can be avoided by a mere change of coordinates. It is the point at which the sectional radius $C(r,t)$ of the collapsing star becomes zero. The path of photons along the light cone boundary is also not a straight path, but a logarithmic curve.
Hence our expression for the time of singularity formation now becomes
\begin{equation}\label{ts}
t_{s}=\frac{1}{C_{3}}\ln\left(\frac{C_{4}\pm\sqrt{\left(C_{4}^{2}+C_{1}C_{2}C_{3}^{2}\right)}}{C_{1}C_{3}}\right),
\end{equation}
which leads to the condition
\begin{equation}\label{tscondition}
-\frac{C_{4}^{2}}{C_{3}^{2}}\le C_{1}C_{2},
\end{equation}
for real values of the logarithm. Moreover, the logarithm needs to have a positive argument.

\subsection{Case II : $D(r)=0$}\label{CaseII}
After putting $D(r)=0$ in the differential equation \eqref{diffeqn} and solving it
for $w$, with the help of GRTensor software package, while considering $w$ to be positive, since it is the temporal part of the sectional radius of the star, we obtain,
\begin{equation}
w=\left[-\frac{d}{c}t^{2}-2C_{5}t+2C_{6}\right]^{\frac{1}{2}},
\end{equation}
where $C_{5}$ and $C_{6}$ are integration constants.
Since the ratio $\frac{d}{c}$ appears in the expression for $w(t)$, which has no radial dependence, we can use the condition $d/c=C_{4}$ in the homogeneous case also, at which, the above expression becomes,
\begin{equation}\label{wdefine}
w=\left[-C_{4}t^{2}-2C_{5}t+2C_{6}\right]^{\frac{1}{2}},
\end{equation}
The significance of this ratio $\frac{d}{c}$ being a constant has been discussed under \textbf{``Case I : $D(r)\ne0$''} of subsection A above.
Solving for the time $t_{s}$ when $w=0$, we get
\begin{equation}\label{ts0case}
t_{s}=\frac{-C_{5}\pm\sqrt{y}}{C_{4}},
\end{equation}
where
\begin{equation}\label{ydefine}
y=C_{5}^{2}+2C_{4}C_{6}.
\end{equation}
From this equation, for a real solution, the following constraint on the constants becomes evident immediately :
\begin{equation}
C_{5}^{2}+2C_{4}C_{6}\ge 0.
\end{equation}
The time derivative of $w$ is given by
\begin{equation}
\dot{w}=\frac{-C_{4}t-C_{5}}{w}.
\end{equation}
To ensure collapse to progress, we require $\dot{w}<0$. Hence at $t=t_{s}$, when $w\rightarrow0$, we see that $\dot{w}\rightarrow -\infty$.
We note that the right-hand side of the field equations for $R_{11}$ and $R_{22}$ involving $\lambda$ are identical, and get cancelled out as we apply the pressure isotropy condition. Hence the parameter $\lambda$ in the $f(R,T)$ function does not appear in the expression for $t_{s}$.

\section{Formation of Apparent Horizon}
The apparent horizon is a boundary between outward directed light rays which bend inwards, and those which move outwards. Several authors have determined the time of formation of apparent horizon for various types of collapse in GR \cite{ChattGhoshJaryal,DebnathNathChak}. Chakrabarti and Banerjee \cite{ChakBanj2016} found apparent horizon formation time for a perfect fluid collapse in $f(R)$ gravity. Sharif and Kausar \cite{Sharif2010} found apparent horizons for spherically symmetric perfect fluid collapse in $f(R)$ theory. Amir and Sattar \cite{Amir2015} found apparent horizons for the collapse of perfect fluid with spherical symmetry in $f(R,T)$ gravity. Abbas and Ahmed \cite{Abbas2019} studied apparent horizon formation for charged perfect fluid collapse in $f(R,T)$ theory.
For the formation of the apparent horizon, we have the requirement
that any outward normal on its boundary must be null. This condition can be
expressed as
\begin{equation}
g^{\mu\nu}C_{,\mu}C_{,\nu}=0,
\end{equation}
which gives us the following condition:
\begin{equation}
\frac{\dot{C}^{2}}{A^{2}}=\frac{C'^{2}}{B^{2}}.
\end{equation}

Using the expressions for $A$, $B$ and $C$ from \eqref{A}, \eqref{B} and \eqref{C}, we arrive at the relation
\begin{equation}\label{deltadefining}
\frac{\dot{w}^{2}}{w^{2}}=\frac{\left(B_{0}+rB_{0}'\right)^{2}A_{0}^{2}}{r^{2}B_{0}^{4}}=\delta^{2}.
\end{equation}

Since, the left hand side is a function of $t$ only, and the right
hand side is a function of $r$ only, so $\delta^{2}$ must be a constant. It is noted that $\delta = 0$ would imply that $w$ would become a constant and hence collapse would not be possible, since the value of $\delta$ should remain unchanged throughout the collapse.
In order to examine the possible values of $\delta$, it is useful to solve the second equation of \eqref{deltadefining} first.

Using the previous expressions for $c$, $d$ and $C_{4}$ from \eqref{cdefine}, \eqref{ddefine} and \eqref{C4define} in \eqref{deltadefining} which follows from the condition for apparent horizon formation, we can
write
\begin{equation}
\frac{\delta^{2}}{C_{4}}=\frac{\left(B_{0}+rB_{0}'\right)^{2}}{B_{0}^{2}}.
\end{equation}

Integrating with respect to $r$, we get
\begin{equation}\label{B0}
B_{0}=C_{7}r^{n}.
\end{equation}
Putting it back in the expression for $\delta^{2}$, we obtain
\begin{equation}\label{A0}
A_{0}=\frac{C_{7}r^{n+1}\delta}{n+1}.
\end{equation}

Here, $C_{7}$ is the integration constant, and $n$ is a real number not equal to -1, so as to keep the metric coefficient $A_{0}$ finite. We also see that $D(r)$ does not vanish with these expressions for $A_{0}$ and $B_{0}$. Hence, these power series solutions are valid only for the case when $D(r)\ne0$. It follows from the condition for apparent horizon, and the form of the metric coefficients, that apparent horizon will be formed only in the case when $D(r)\ne0$.

For collapse to proceed, it is necessary that the condition $\dot{w}<0$ be satisfied. Using the form of $w$ and $\dot{w}$ obtained from equation \eqref{w}, where $D(r)\ne0$, this leads us to the following constraint :
\begin{equation}\label{collapseconstraint}
e^{-2\delta t}\left(\delta+6C_{2}e^{-2\delta t}\right)-2C_{1}<0.
\end{equation}

\bigskip

Using the power series forms of $A_{0}(r)$ and $B_{0}(r)$ from equations \eqref{A0} and \eqref{B0}, we find that
\begin{equation}
C_{3}=H=2\delta ,
\end{equation}
and
\begin{equation}\label{C4}
C_{4}=\frac{A_{0}^{2}}{r^{2}B_{0}^{2}}=\frac{\delta^{2}}{\left(n+1\right)^{2}},
\end{equation}
which is a positive quantity for real values of $\delta$ and $n$. We choose $n=1$ here onwards for our purpose.

The expression for the time of formation of singularity given by equation \eqref{ts} is independent of the radial coordinate $r$. So it is expected that all the shells of the collapsing spherical matter ball will collapse to the final singularity at the same time $t_{s}$. In such a case, the physical radius of the outer shells should decrease at a faster rate compared to the inner shells, without however crossing each other (thus avoiding any shell-crossing), so that all the shells reach the final singularity at the same time.

Equation \eqref{ts} can be rewritten in the form
\begin{equation}\label{tsrewrite}
t_{s}=\frac{1}{2\delta}\ln\left(\frac{\delta^{2}\pm4\sqrt{z}}{8\delta C_{1}}\right),
\end{equation}
where,
\begin{equation}\label{zdefine}
z=C_{4}^{2}+C_{1}C_{2}C_{3}^{2}=\frac{\delta^{4}}{16}+4\delta^{2}C_{1}C_{2}.
\end{equation}
For real values of the discriminant, $z$ must be positive definite.
Using the forms of $w$ and $\dot{w}$ obtained from equation \eqref{w} where $D(r)\ne0$, and using them in the condition for apparent horizon, that is,
\begin{equation}
\dot{w}^{2}=\delta^{2}w^{2} ,
\end{equation}
and solving for the apparent horizon formation time, and utilising equation \eqref{ts}, we have
\begin{equation}\label{tah1}
t_{ah_{1}}-t_{s}=\frac{1}{2\delta}\ln\left(\frac{-64C_{1}C_{2}}{\delta\left(\delta\pm\sqrt{\delta^{2} +64C_{1}C_{2}}\right)}\right) ,
\end{equation}
and
\begin{equation}\label{tah2}
t_{ah_{2}}-t_{s}=\frac{1}{2\delta}\ln\left(\frac{3\delta\pm\sqrt{9\delta^{2} +512C_{1}C_{2}}}{2\delta\pm\sqrt{4\delta^{2}+256C_{1}C_{2}}}\right).
\end{equation}
When expressed in terms of $z$ and $\delta$, we have,
\begin{equation}\label{tah1altform}
t_{ah_{1}}-t_{s}=\frac{1}{2\delta}\ln\left(\frac{\delta^{4}-16z}{\delta^{2}\left(\delta^{2}\pm4\sqrt{z}\right)}\right),
\end{equation}
and
\begin{equation}\label{tah2altform}
t_{ah_{2}}-t_{s}=\frac{1}{2\delta}\ln\left(\frac{3\delta^{2}\pm\sqrt{\delta^{4} +128z}}{2\left(\delta^{2}\pm4\sqrt{z}\right)}\right).
\end{equation}

\subsection*{Conditions for formation of Black Hole}
Joshi, Goswami and Dadhich \cite{JoshiGosDadhPRD04} showed that in a case like above, where the time of formation of singularity is not dependent on the radial coordinate, no naked singularity can result. So it is necessary to examine the constraints on the parameters $\delta$, $C_{1}$ and $C_{2}$ for black hole formation to be possible.

If a black hole is to be formed, it is necessary that $t_{ah}-t_{s}<0$. This would imply that $t_{ah} < t_{s}$, which shows that the apparent horizon forms before the singularity. In other words, the singularity is hidden behind the apparent horizon, and is a black hole.

Let us examine the constraints on the possible cases for this inequality to hold.

\subsection*{Case I : the condition $z>0$}

\begin{enumerate}
\item For the expression $t_{ah_{1}}-t_{s}$ from equation \eqref{tah1altform} to be negative for black hole formation, the following constraints must be obeyed :
\begin{table}[!h]
\begin{center}
\caption{Constraints on $C_{1}$ and $C_{2}$ from $t_{ah_{1}}-t_{s}<0$ for Black Hole Formation}
\label{Table I}
\begin{tabular}{|c|c|c|c|}
\hline
& & &\\
& \textbf{Sign of $\delta$} & \textbf{Sign chosen from $\pm$} & \textbf{Condition for Black Hole Formation} \\
& & \textbf{from denominator} &\\
\hline
\textbf{1.} & $\delta>0$ & $-$ & Black Hole formation is not possible \\
& & &\\
\hline
\textbf{2.} & $\delta>0$ & $+$ & $-\left(\delta^{2}/64\right)<C_{1}C_{2}<0$ \\
& & &\\
\hline
\textbf{3.} & $\delta<0$ & $+$ & $-\left(\delta^{2}/64\right)<C_{1}C_{2}<0$ or $C_{1}C_{2}>0$ \\
& & &\\
\hline
\textbf{4.} & $\delta<0$ & $-$ & Black Hole formation is not possible \\
& & &\\
\hline
\end{tabular}
\end{center}
\end{table}

\item For the expression $t_{ah_{2}}-t_{s}$ from equation \eqref{tah2altform} to be negative for black hole formation, the following constraints are to be obeyed :
\begin{table}[!h]
\begin{center}
\caption{Constraints on $C_{1}$ and $C_{2}$ from $t_{ah_{2}}-t_{s}<0$ for Black Hole Formation}
\label{Table II}
\begin{tabular}{|c|c|c|c|c|}
\hline
& & & &\\
& \textbf{Sign of $\delta$} & \textbf{Sign chosen from $\pm$} & \textbf{Sign chosen from $\pm$} & \textbf{Condition for } \\
& & \textbf{in numerator} & \textbf{denominator} & \textbf{Black Hole Formation}\\
\hline
\textbf{1.} & $\delta>0$ & $+$ & $+$ & Black Hole formation is not possible \\
& & & &\\
\hline
\textbf{2.} & $\delta>0$ & $-$ & $-$ & Black Hole formation is not possible \\
& & & &\\
\hline
\textbf{3.} & $\delta>0$ & $+$ & $-$ & Black Hole formation is not possible \\
& & & &\\
\hline
\textbf{4.} & $\delta>0$ & $-$ & $+$ & $C_{1}C_{2}>-\left(\delta^{2}/64\right)$ \\
& & & &\\
\hline
\textbf{5.} & $\delta<0$ & $+$ & $+$ & $-\left(\delta^{2}/64\right)<C_{1}C_{2}<0$, or $C_{1}C_{2}>0$ \\
& & & &\\
\hline
\textbf{6.} & $\delta<0$ & $-$ & $-$ &  $-\left(\delta^{2}/64\right)<C_{1}C_{2}$\\
& & & &\\
\hline
\textbf{7.} & $\delta<0$ & $+$ & $-$ & Black Hole formation is not possible \\
& & & &\\
\hline
\textbf{8.} & $\delta<0$ & $-$ & $+$ & $-\left(\delta^{2}/64\right)<C_{1}C_{2}<0$ \\
& & & &\\
\hline
\end{tabular}
\end{center}
\end{table}
\end{enumerate}

\newpage

\subsection*{Case II : the condition $z=0$}
\begin{enumerate}
\item For $z=0$, the expression $t_{ah_{1}}-t_{s}$ from equation \eqref{tah1altform} becomes zero, and black hole formation is not possible.
\item For the expression $t_{ah_{2}}-t_{s}$ from equation \eqref{tah2altform} to be negative for black hole formation, the following constraints are to be obeyed :
\begin{table}[!h]
\begin{center}
\caption{Conditions for Black Hole Formation from $t_{ah_{2}}-t_{s}<0$}
\label{Table III}
\begin{tabular}{|c|c|c|}
\hline
& &  \\
& \textbf{Value of $t_{ah_{2}}-t_{s}$} & \textbf{Condition for Black Hole Formation} \\
& & \\
\hline
\textbf{1.} & $0$ & Black Hole formation is not possible \\
& & \\
\hline
\textbf{2.} & $\frac{1}{2\delta}\ln 2$ & $\delta<0$ \\
& & \\
\hline
\end{tabular}
\end{center}
\end{table}
\item There is an additional constraint for the case $z=0$ : \textbf{$\delta$ and $C_{1}$ should have the same sign}.
\end{enumerate}

\section{Summary and Discussions}

To summarize, we have started with a generalized Vaidya exterior metric, considering the exterior spacetime to the collapsing matter
to be filled with a combination of Type-I and Type-II matter fluid. We
investigate the non-adiabatic collapse of an isotropic matter ball involving heat
flux. The form of the temoral dependence of the physical radius from the pressure isotropy condition by utilising the field equations in $f(R,T)$ theory of gravity for a collapsing matter involving heat flux had not been previously found in any literature. Proceeding with the collapse formalism, the field equations
are formed, the $f(R,T)$ junction conditions are applied with the generalised Vaidya exterior
solution. Finally, for determining the time of formation of the final singularity at the end of the collapse, the metric coefficients
for the interior spacetime, $A(r,t)$ and $B(r,t)$ are assumed to
be functions of the radial coordinate, and, in case of $C(r,t)$ which
represents the physical radius of the collapsing fluid, to be separable
in the spatial and temporal coordinates. Using these forms of the metric parameters, a differential
equation in the temporal function $w$ is obtained by utilising
the condition of pressure isotropy. By requiring $w(t)=0$, the time of the singularity
formation $t_{s}$ is obtained with the necessary restrictions imposed
on the integration constants, for two cases : i) when $D(r)\ne0$, and ii) when $D(r)=0$.
A power series solution in the radial
coordinates is obtained for $A_{0}(r)$ and $B_{0}(r)$, from the condition for apparent horizon formation and it is seen that $D(r)$ is non-vanishing in that case. The time of formation of apparent
horizon, $t_{ah}$ is also obtained. The difference between $t_{ah}$ and $t_{s}$ for $D(r)\ne0$ is calculated, and the constraints on the parameters $\delta$, $C_{1}$ and $C_{2}$ are examined for the final singularity to be a black hole.

The following observations can be made in the context of this work
:
\begin{enumerate}
\item We have shown in our analysis, just as Rosa had considered a priori in \cite{RosaPRD2021} while formulating the $f(R,T)$ junction conditions, that the interior matter Lagrangian must match the exterior matter Lagrangian at the boundary of the collapsing matter. Also, the derivatives of the matter Lagrangian with respect to the coordinates of their respective spacetimes, should be continuous across the boundary. It is worth noting that this condition arises from the matching of the Ricci scalars and the traces of the energy momentum tensors of the interior and the exterior regions across the boundary of the collapsing matter, which are the junction conditions for $f(R,T)$ theory. These conditions need not be satisfied in GR, as a result of which it is not necessary in GR for the matter Lagrangians and their derivatives to be matched at the boundary.
It is seen that the absence of heat flux (perfect fluid case), and
utilising equations \eqref{A}, \eqref{B} and \eqref{C} would give us a relation between $A_{0}(r)$
and $B_{0}(r)$ from equation \eqref{field01}. Their ratio would still be linear in
the radial coordinate.
\item If the pressure was anisotropic, then the pressure in equation \eqref{pqrel}
would be the radial pressure, since the field equation for $R_{11}$, which is the radial component of the Ricci tensor,
was used to arrive at it.
\item In case of pressure anisotropy, we would need to find a different
way to arrive at the solution for the temporal function $w$.
In that case, equating the expression for $q$ from equation \eqref{pqrel}
and equation \eqref{field01} and using the equations \eqref{A}, \eqref{B} and \eqref{C} would have provided
a differential equation for $w$ with additional terms involving $w$ and its time derivative.
This would have resulted in a more complicated relation to work with.
\item Presence of shear would have further added terms involving $w$
and its time derivative to the differential equation, since it would
have appeared in the field equation for $R_{11}$ and consequently
in equation \eqref{pqrel}. Hence the corresponding differential equation would become more
complicated if effects like anisotropy and shear are involved in the collapse.
\item The parameter $\lambda$ has no effect on either the singularity formation or the apparent horizon formation : the former because it gets cancelled out as the pressure isotropy condition is applied, and the latter because the condition for the formation of apparent horizon remains the same in $f(R,T)$ theory, as in GR.
\item We see that for a black hole to form as the final state of the collapse, $\delta$, $C_{1}$ and $C_{2}$ get constrained in different ways for the possible expressions of $t_{ah}-t_{s}$, obtained from the condition for apparent horizon, for non-vanishing $D(r)$, both when $z$ is zero and non-zero.
\item Tables I and II enlist the constraints that must be satisfied for a black hole to form when $z>0$ for two possible expressions of $t_{ah}-t_{s}$, while Table III shows the constraints required for black hole formation for the $z=0$ case.
\item The chosen configuration of the collapsing matter does not involve shear viscosity. It has been reasoned both in GR \cite{JoshiGoswamiDadhicharxiv2004}, as well as in $f(R,T)$ theory \cite{AzmatZubairNoureenIJMPD2017} that when shearing effects come into play, the matter concentration will be resisted by these forces. As a result, formation of a trapped surface or apparent horizon will be delayed, and the probability of a naked singularity as the final state of the collapse will be greater. It is also concluded in \cite{JoshiGoswamiDadhicharxiv2004}, that the absence of shear would necessarily have a black hole as the final state of collapse. Since the retardation of the collapse and delay of formation of apparent horizon is a result of the shearing effect and not specifically dependent on the gravity theory chosen, it is reasonable to predict that in our case, where the collapse takes place in $f(R,T)$ gravity, and in absence of shear, the end result must necessarily be a black hole. Hence the constraints obtained on the parameters $\delta$, $C_{1}$ and $C_{2}$, will have to be adhered to.
\item In this paper, we have considered the gravitational collapse of a spherical non-rotating body where a spherical black hole will be formed as the final state of the collapse in the cases where black hole formation is possible \cite{FrolovNovikov1998}. Since the collapsing matter is radiating heat in the radial direction, mass-energy content of the collapsing body will decrease with time. In a previous paper of ours \cite{GuhaGhosh2021}, equation (40) provides an expression for the temporal variation of the mass-energy content of a collapsing matter in $f(R,T)$ gravity. Adjustments can be made in that expression which would fit our case for an $f(R,T)$ function linear in both $R$ and $T$, pressure isotropy and absence of shear and free-streaming radiation. The adjusted equation reads

\begin{equation}\label{tmv}
D_T M=-\frac{C^2}{2f_R}\left[U\left\{(1+f_T)p+\rho f_T+\frac{1}{2}\left(f-Rf_{R}\right)\right\}
+H(1+f_T)q\right].
\end{equation}

All the terms of the expression are explained in \cite{GuhaGhosh2021}. Since the time of formation of apparent horizon has been calculated, the amount of mass-energy content left in the collapsing sphere just after the apparent horizon forms, may be determined via an integration over time from $0$ to $t_{ah}$ provided the functional forms of all the terms are explicitly known. This may give us an indication of the mass of the black hole formed. As stated in \cite{HawkingEllis, Faraoni2013}, in the case of spherical collapse of uncharged matter, the event horizon and apparent horizon coincide as the collapsed matter reaches the final static state. Hence the same will be true in our case.
\end{enumerate}
In this context, we want to mention that instead of considering a linear function of $R$ and $T$, if we consider some other functional dependence of $R$, for example, a generalised function like $f(R,T)=\alpha R^{n}+\beta T$ where $\alpha$ and $\beta$ are constants, terms involving the dark source terms $D_{\mu\nu}$ and $f_{R}$ will appear in the pressure isotropy condition \eqref{pressureisotropy} and the resulting differential equation \eqref{diffeqn} will become still more complicated and difficult to solve. These terms would arise entirely from the geometry of the spacetime and the modified gravity function. It is only because we have chosen a linear $f(R,T)$ function, that $f_{R}$ becomes unity and all the components of $D_{\mu\nu}$ vanish. Consequently, the pressure isotropy condition gives rise to a relatively less complicated differential equation. The contribution from modified gravity will be present in equation \eqref{diffeqn} for other functional forms of $f(R,T)$. Also, equation \eqref{tmv}, which describes the temporal variation of the mass-energy content of the collapsing matter, already has the terms $f_{R}$ and $f_{T}$ which comes from the modified gravity. Although $f_{R}$ becomes unity in our case, $f_{T}$ takes the value of the parameter $\lambda$. In the manner of the prescription provided in \cite{CanateJaimeSalgado}, extending the method to $f(R,T)$ gravity, black hole solutions may be obtained from the $f(R,T)$ field equations of the generalized Vaidya exterior spacetime, by integrating over the radial coordinate from the black hole horizon to the asymptotic region. Determining the black hole area from the horizon radius and consequently the surface gravity may be possible following the ideas about black hole thermodynamics in modified gravity \cite{HazarikaPhukon2024, DuttaMukhopadhyay2024}.

There is scope of further investigation as to what happens in presence of shear viscosity and pressure anisotropy in the collapsing star, which may be taken up in future.

\section*{Acknowledgements}
The authors are thankful to anonymous reviewers for their comments and helpful suggestions. The authors are also grateful to Dr. Sunil Maharaj for his comments and suggestions on this work. SG thanks IUCAA, India for an associateship. A portion of this work was done in IUCAA, India under the associateship program. The authors are thankful for the warm hospitality and the facilities of work available in IUCAA.

\section*{Appendix A}

This is the elaborate method of obtaining equations \eqref{Hconstant} and \eqref{C4define}, by showing why the ratio $\frac{d(r)}{c(r)}$ is a constant both for the case when $D(r)\ne0$ and when $D(r)=0$. In each case we start with the solution for the function $w(t)$.

For $D(r)\ne0$, the solution of $w(t)$ is given by \eqref{w}:

\begin{equation}\label{wappendix}
w(t)=\pm\frac{1}{\sqrt{2Dk^{3}}}\left(C_{1}Hck^{2}-2dk-C_{2}Hc\right)^{1/2},
\end{equation}
where $C_{1}$ and $C_{2}$ are integration constants,
\begin{equation}\label{c}
c(r)=\frac{1}{A_{0}^{2}},
\end{equation}
\begin{equation}\label{d}
d(r)=\frac{1}{r^{2}B_{0}^{2}},
\end{equation}
\begin{equation}\label{H}
H(r)=\sqrt{\frac{2D}{c}},
\end{equation}
and,
\begin{equation}\label{k}
k=\textrm{exp}(Ht).
\end{equation}
From \eqref{wappendix} and using \eqref{H}, we get,
\begin{equation}\label{wsq}
w^{2}=\frac{C_{1}}{Hk}-\frac{d}{Dk^{2}}-\frac{C_{2}}{Hk^{3}}.
\end{equation}
Since the left-hand side of the above equation is a function of $t$ only, the right-hand side must also be a function of $t$ and have no dependence on $r$.
There are 3 additive terms on the right hand side. So each of them must individually be a function of $t$.
Examining the first term on the right-hand side of \eqref{wsq}, we see that $C_{1}$ must be a constant.
So, $Hk$ must be independent of $r$. In other words, by requiring its derivative with respect to $r$ to vanish, and using \eqref{k}, we have
\begin{equation}\label{conditionforH}
H'\left(1+Ht\right)e^{Ht}=0.
\end{equation}
where prime denotes derivative with respect to $r$.
Now, $e^{Ht}\ne0$ for real values of $H(r)$ and $t$.
If we consider $1+Ht=0$, it implies,
\begin{equation}
H=-\frac{1}{t},
\end{equation}
which is a contradiction, since, $H$ is a function of $r$.
So the only possibility for \eqref{conditionforH} to be satisfied, is that
\begin{equation}
H'=0,
\end{equation}
which implies that $H$ must be a constant, and thus equation \eqref{Hconstant} is obtained.
From \eqref{H}, we can also see that the ratio
\begin{equation}\label{Dcratio}
\frac{D(r)}{c(r)} = \textrm{constant}=m_{1}.
\end{equation}
Since $H$ is a constant, we also have from \eqref{k} that $k$ is a function of $t$ only.
Taking the second term of equation \eqref{wsq}, since $k=k(t)$, it follows that $d/D$ should also be a function of $t$ alone. But since, $d$ and $D$ are functions of $r$ without any temporal dependence, the only way this is possible is if their ratio is a constant. Hence we have,
\begin{equation}\label{dDratio}
\frac{d(r)}{D(r)}=\textrm{constant}=m_{2}.
\end{equation}
Multiplying \eqref{Dcratio} and \eqref{dDratio} and using the fact that $D(r)\ne0$, we see that
\begin{equation}\label{dcratio}
\frac{d(r)}{c(r)}=m_{1}m_{2}=\textrm{constant}=C_{4}.
\end{equation}
Thus equation \eqref{C4define} is obtained.

\bigskip

\underline{\textbf{Comments:}}

For $D(r)=0$, the solution of $w(t)$ is given by
\begin{equation}
w=\left[-\frac{d}{c}t^{2}-2C_{5}t+2C_{6}\right]^{\frac{1}{2}},
\end{equation}
where $C_{5}$ and $C_{6}$ are integration constants, $c$ and $d$ are as defined in \eqref{cdefine} and \eqref{ddefine}.
Squaring both sides,
\begin{equation}\label{wsquare}
w^{2}=-\frac{d}{c}t^{2}-2C_{5}t+2C_{6}.
\end{equation}
Since the left-hand side of \eqref{wsquare} is a function of $t$ alone, the right-hand side must also be a function of $t$ and independent of $r$. It is obvious from the fact that $C_{5}$ and $C_{6}$ are constants, that the second and the third terms are function of $t$ only, and a constant, respectively.
For the term $\frac{d}{c}t^{2}$ to be independent of $r$, the derivative of $\frac{d}{c}$ with respect to $r$ must vanish.
But since both $d(r)$ and $c(r)$ are functions of $r$, the only way the radial derivative of this ratio would vanish is if $\frac{d}{c}$ is a constant.
Hence, for both vanishing and non-vanishing $D(r)$, the ratio $\frac{d(r)}{c(r)}$ must be a constant.

\end{document}